%% file: main.tex
\documentclass[conference]{IEEEtran}
\IEEEoverridecommandlockouts
\usepackage{hyperref}
\usepackage{mathtools}
\usepackage{tcolorbox}
\usepackage{color}
\usepackage{theorem}
\usepackage{times,amsmath,epsfig}
\usepackage{amssymb}
\usepackage{subfigure}
\usepackage{cite}
\usepackage{cases}
\usepackage{siunitx}
\usepackage{algorithmic}
\usepackage{relsize}
\usepackage{algorithm}
\usepackage{graphicx}
\usepackage{bbm}
\usepackage{bm}
\input{my_sections.tex}
\usepackage{tikz}
\usetikzlibrary{shapes,arrows}
\input{mysymbol.sty}
\input{figures/tikz_styles}

\newtheorem{remark}{\bf Remark}

\begin{document}

\title{Learning Non-myopic Power Allocation in Constrained Scenarios}
\author{Arindam Chowdhury$^\star$, Santiago Paternain$^\#$, Gunjan Verma$^\dag$, Ananthram Swami$^\dag$, and Santiago Segarra$^\star$ 
        \\ \\
        $^\star$Dept. of ECE, Rice University  \hspace{1cm} $^\#$Dept. of ECSE, Rensselaer Polytechnic Institute \\ $^\dag$DEVCOM Army Research Laboratory
\thanks{Research was sponsored by the Army Research Office and was accomplished under         Cooperative Agreement Number W911NF-19-2-0269. 
        The views and conclusions contained in this document are those of the authors and should not be interpreted as representing the official policies, either expressed or implied, of the Army Research Office or the U.S. Government. 
        The U.S. Government is authorized to reproduce and distribute reprints for Government purposes notwithstanding any copyright notation herein. 
        \newline
        Emails: \{ac131, segarra\}@rice.edu, paters@rpi.edu, \{gunjan.verma, ananthram.swami\}.civ@army.mil.}} 

\maketitle

\begin{abstract}
We propose a learning-based framework for efficient power allocation in ad hoc interference networks under episodic constraints.
The problem of optimal power allocation -- for maximizing a given network utility metric -- under instantaneous constraints has recently gained significant popularity. Several learnable algorithms have been proposed to obtain fast, effective, and near-optimal performance.
However, a more realistic scenario arises when the utility metric has to be optimized for an entire episode under time-coupled constraints.
In this case, the instantaneous power needs to be regulated so that the given utility can be optimized over an entire sequence of wireless network realizations while satisfying the constraint at all times.
Solving each instance independently will be myopic as the long-term constraint cannot modulate such a solution.
Instead, we frame this as a constrained and sequential decision-making problem, and employ an actor-critic algorithm to obtain the constraint-aware power allocation at each step.
We present experimental analyses to illustrate the effectiveness of our method in terms of superior episodic network-utility performance and its efficiency in terms of time and computational complexity.
\end{abstract}


\section{Introduction}\label{S:intro}

Power allocation for interference management in wireless networks~\cite{shannon1948mathematical} is essential for satisfying high quality-of-service (QoS) requirements in modern communication systems~\cite{liang2019deep}. 
Mathematically, it can be formulated as the problem of optimizing a certain system-level utility function (such as sum-rate or harmonic-rate) subject to instantaneous resource budget constraints. However, such a problem formulation is NP-hard~\cite{luo_2008_dynamic}, and a classical approximate solution involves reformulating it in the form of a surrogate tri-convex objective with instantaneous constraints and solving it via an iterative block-coordinate-descent based approach termed WMMSE~\cite{shi2011iteratively}.
Since the advent of deep learning, several \textit{data-driven} alternatives have been proposed~\cite{eisen2019learning,shen2019graph,hu2020iterative,pellaco2021deep,chowdhury2023deep,schynol2023coordinated,li2022graph}, that try to address multiple drawbacks in the WMMSE algorithm, including computational and time complexity~\cite{chowdhury2021unfolding}. 
However, none of these algorithms are capable of handling time-dependent constraints. 
For example, episodic sum-rate maximization under battery constraints is a crucial problem, especially in scenarios wherein the available battery is limited and channel conditions vary over time.
The key challenge here is to selectively allocate power under favorable channel conditions only, such that the available battery can be preserved for a longer duration in an episode, resulting in a \textit{non-myopic} (non-greedy) power allocation method.

The power allocation process at successive time steps, based on the current channel conditions and battery level, can be modeled as a sequential decision-making problem under constraints. 
Further, considering the fact that the instantaneous power allocation depends only on the last battery state and not the entire history, we frame this as a Markov decision process~\cite{sutton2018reinforcement}.
Recently, the paradigm of constrained reinforcement learning (CRL)~\cite{altman2021constrained,junges2016safety,paternain2019constrained,paternain2019safe} has become an extremely popular data-driven framework to tackle episodic optimization problems with time-coupled constraints.
In this work, we combine the risk-aware reward function formulation~\cite{geibel2005risk} with constrained policy~\cite{paternain2019safe} to develop a hierarchical framework for learning \textit{non-myopic power allocation} (NMPA) in wireless networks under battery constraints.

\noindent \textbf{Notation}: The entries of a matrix $\mathbf{X}$ and a vector $\mathbf{x}$ are denoted by $[\mathbf{X}]_{ij}$ and $[\mathbf{x}]_i$.
The all-zeros and all-ones vectors are denoted by $\mathbf{0}$ and $\mathbf{1}$.
$[\cdot]_+$ represents a $\max(\cdot,0)$ operation.
$\mathbb{E}(\cdot)$ is the expectation operator. The zero-mean normal distribution of variance $\sigma^2$ is denoted by $\ccalN(0, \sigma^2)$.

\section{System model and problem formulation}\label{S:Modeling}

We start by laying out a procedure for constructing a finite, time-indexed set of wireless single-hop ad-hoc interference networks having $M$ distinct single-antenna transceiver pairs. 
Time is slotted into uniform intervals indexed by $t$. 
At each time step $t$, a network realization is sampled randomly from a joint distribution over network topologies and fading conditions. 
In each sample, the transmitters are denoted by $i$ and their unique intended receivers are denoted by $r(i)$ for $i \in \{1, \ldots, M\}$. 
The transmitters are dropped uniformly at random as $\mathbf{l}_i \in [-S, S]^2$, $S$ being user-defined, and their corresponding receivers are dropped uniformly at random as $\mathbf{o}_i \in [\mathbf{l}_i - \frac{R_i}{\sqrt{2}}, \mathbf{l}_i + \frac{R_i}{\sqrt{2}}]^2$, where $R_i$ is the range of transmitter $i$. 
The $i$-th transmitter can only communicate with its corresponding receiver $r(i)$. 
It can, however, interfere with receivers $r(j)$, for all $j\neq i$, that lie within its range.
We represent the set of all transmitters interfering with a receiver $r(i)$ as $\mathcal{E}_{r(i)}$.
For simplicity, we assume that all transmitters have equal range i.e $R_i = R$ for all $i$. 
Further, we simulate fading effects in the sample by incorporating small-scale Rayleigh fading as well as large-scale path loss. 

An arbitrary channel coefficient corresponding to the network realization at time $t$ is given by $h(t)$. 
Now, denoting the signal transmitted by $i$ at time $t$ as $x_i(t) \in \mathbb{R}$, and assuming a linear transmission model, the received signal $y_i(t) \in \mathbb{R}$ at $r(i)$ is given by 
%
\begin{equation}\label{E:trans_model}
     y_i(t) = h_{ii}(t)x_i(t) + \sum_{j\in \mathcal{E}_{r(i)}} h_{ij}(t)x_j(t) + n_i(t),
\end{equation}
where $h_{ii}(t) \in \mathbb{R}$ is the channel between the $i$-th transceiver pair at time $t$, $h_{ij}(t) \in \mathbb{R}$ for $j \in \mathcal{E}_i$ represents the interference between transmitter $j$ and receiver $r(i)$, and $n_i(t) \sim \ccalN(0, \sigma_N^2)$ represents the additive channel noise. 
The time-varying channel states $h_{ij}(t)$ are consolidated in a channel-state information (CSI) matrix $\mathbf{H}_t \in \reals^{M \times M}$ where $[\mathbf{H}_t]_{ij} = h_{ij}(t)$.
We define an \textit{episode} of duration $T$ as the sequence of i.i.d. CSI matrices $\{\mathbf{H}_1,\dots,\mathbf{H}_T\}$.

The instantaneous data rate $c_i$ achievable at receiver $r(i)$ at time $t$ is given by,
\begin{equation}\label{E:data_rate}
    c_i(\bbp_t, \bbH_t) = \log_2 \left( 1 + \frac{ [\mathbf{H}_t]_{ii}^2 [\mathbf{p}_t]_i}{\sigma_N^2 + \sum_{j\in \mathcal{E}_{r(i)}} [\mathbf{H_t}]_{ij}^2 [\mathbf{p}_t]_j} \right),
\end{equation}
where $\mathbf{p}_t \ge \mathbf{0}$ 
is the allocated power at time $t$.
In the absence of episodic constraints, which are constraints coupled across time, optimal power allocation for a given CSI $\mathbf{H}_t$ at any time $t$, can be obtained by solving the following sum-rate maximization problem:
\begin{align}
    \max_{\mathbf{p}} \,\, \sum_{i=1}^M c_i(\mathbf{p}, \mathbf{H}_t)~\label{E:optimization_problem} \\   
    \text{s.t.} \,\, \mathbf{p} \in [0, P_{\max}]^M~\label{Eq:cons} 
\end{align}
where $P_{\max} \in \mathbb{R}$ denotes the maximum available power at every transmitter. 

Several solution models for~\eqref{E:optimization_problem}-\eqref{Eq:cons} exist, including iterative block-coordinate descent based optimizers~\cite{shi2011iteratively}, neural network based approaches~\cite{eisen2019learning,shen2019graph} as well as a class of unfolding~\cite{monga2019algorithm} based hybrid methods~\cite{hu2020iterative,pellaco2021deep} combining both. Recently, a GNN-based hybrid model UWMMSE~\cite{chowdhury2021unfolding} was proposed, that leverages the underlying graph structure of wireless networks captured implicitly in $\mathbf{H}$ to allocate power efficiently. 

However, none of these methods is suited to handle episodic constraints, which are far more challenging. For example, under limited battery availability, the task of optimal power allocation would ideally involve transmitting optimally at time instants wherein the channels are ``good'', i.e, the achievable sum-rate is high, while intelligently preserving battery under poor channel conditions so that the overall episodic sum-rate can be maximized.  

To formalize the aforementioned problem, let $\mathbf{b}_t \in \mathbb{R}^M$ denote the battery levels of all transmitters at time $t$. An initial battery budget is given by $\mathbf{b}_0$, where $\max_i[\mathbf{b}_0]_i = B_{\max}$ is user-defined and $\min_{i}{[\mathbf{b}_0]_i}\gg P_{\max}$.
The battery dynamics can be modeled as
\begin{align}
[\mathbf{b}_t]_i = [\mathbf{b}_{t-1}]_i - [\mathbf{p}_{t}]_i -\alpha \mathbbm{1}([\mathbf{p}_{t}]_i > 0)\text{ for all } t,i.~\label{Eq:bat}    
\end{align}
Here, $\alpha$ represents a fixed cost of transmission. 
Note that the batteries for all transmitters evolve non-increasingly over time, effectively meaning that the spent batteries cannot be recharged within a given episode.

Further, by denoting a power allocation function as $f$, such that $\mathbf{p}_t = f(\mathbf{b}_{t-1},\mathbf{H}_t)$, the problem of interest can be defined as:
\begin{align}
    \max_{f \in \mathcal{F}}&\,\,\mathbb{E}_{\substack{{\{\mathbf{H}_1,\dots,\mathbf{H}_T\} \sim \mathcal{H}} \\ \mathbf{b}_0\sim \mathcal{B}}}\Bigg[ \frac{1}{T}\sum_{t=1}^T \sum_{i=1}^M c_{i}(\mathbf{p}_t, \mathbf{H}_t) \Bigg]~\label{Eq:opt}\\   
    &\text{s.t.} \,\, \mathbf{p}_t \leq P_{\max} \mathbf{1} \text{ for all } t~\label{Eq:cons1} \\
    &\quad \,\, \ \mathbf{b}_t \geq \mathbf{0} \text{ for all } t,~\label{Eq:cons2}
\end{align}
where $\mathcal{F}$ is a suitably chosen space of continuous functions, $\mathcal{H}$ is an unknown stationary distribution over CSI matrices and $\mathcal{B}$ is an unknown distribution over battery budgets.

A careful analysis of~\eqref{Eq:bat} and~\eqref{Eq:cons2} reveals that the main purpose of the battery constraint is to restrict $f$ from allocating power ever again within an episode to transmitters that fully deplete their battery, given the non-increasing battery dynamics.
This implicitly forces $f$ to utilize the available battery budget judiciously in a given episode.
It is important to note here that solving the optimization problem~\eqref{Eq:opt}-\eqref{Eq:cons1}, without satisfying~\eqref{Eq:cons2}, is equivalent to solving~\eqref{E:optimization_problem}-\eqref{Eq:cons} over multiple time-steps $t=\{1,\dots,T\}$ independently. 
We define this method as the \textit{myopic} (greedy) power allocation model.  
It is clearly ineffective as it becomes inapplicable once~\eqref{Eq:cons2} is violated. 
For example, if batteries deplete at any $t<T$, then the achievable sum-rate for the remaining CSI matrices $\{\mathbf{H}_{t+1},\dots,\mathbf{H}_T\}$ will be null.
The main goal of this work, therefore, is the development of a \textit{non-myopic} power allocation model to solve~\eqref{Eq:opt} under the time-coupling introduced by~\eqref{Eq:bat} and the corresponding constraint~\eqref{Eq:cons2}, while satisfying~\eqref{Eq:cons1} at all $t$.

\section{Proposed Method}\label{S:proposed}

The optimization problem~(\ref{Eq:opt}-\ref{Eq:cons1}) is non-convex and NP-hard~\cite{luo2008dynamic}, even for a deterministic episode (a fixed set of CSI matrices) without any episodic constraints. 
The stochasticity in~\eqref{Eq:opt} and the battery constraint~\eqref{Eq:cons2} further add to the complexity of the problem.
Moreover, the optimization in~\eqref{Eq:opt} is over an infinite-dimensional space $\mathcal{F}$ of continuous functions.

To address these challenges, we decouple the instantaneous power allocation objective from the battery constraint at all time steps.
Since the instantaneous problem has already been solved effectively~\cite{hu2020iterative,pellaco2021deep,chowdhury2021unfolding}, we employ a suitable solver to construct a hierarchical model at two levels. 
The lower level generates an approximate solution $\bar{\mathbf{p}}_t$ of the optimal instantaneous power allocation at each time-step $t$ based purely on the corresponding $\mathbf{H}_t$ without considering the state of the battery $\mathbf{b}_t$. The upper level then modulates $\bar{\mathbf{p}}_t$ such that the battery constraint is satisfied over the entire episode. Denoting the lower-level solver as $g$ and the upper-level solver as $\pi$, we propose the following solution model:
\begin{align}
    \mathbf{p}_t &= \pi(\mathbf{b}_{t-1}, \mathbf{H}_t; \bar{\mathbf{p}}_t) \text{ for all } t\\
    \text{where,}\,\, \bar{\mathbf{p}}_t &= g(\mathbf{H}_t) \text{ for all } t,
\end{align}
and $\mathbf{b}_t$ evolves as per~\eqref{Eq:bat}. 
%
\begin{figure}[!t]
	\centering
	\includegraphics[width=\linewidth,height=0.45\linewidth]{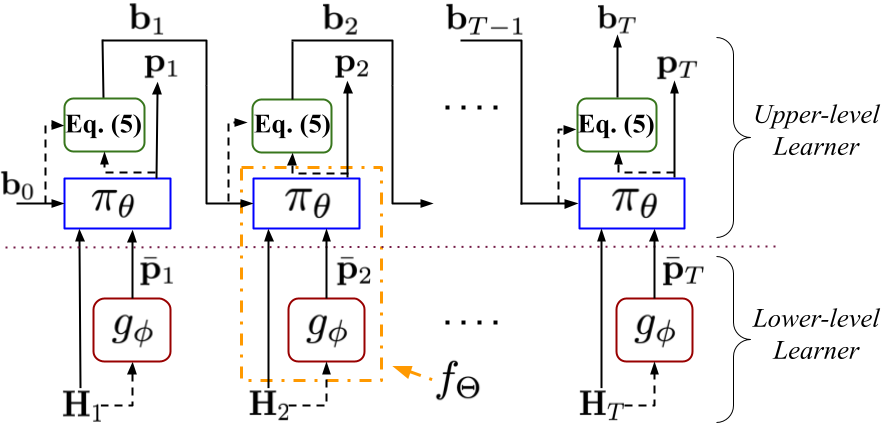}
	\vspace{-0.4cm}
	\caption{\small {Block diagram of the proposed model over $T$ time-steps.}}
	\label{F:block}
\end{figure}
%
Further, due to the relative ease of sampling from the unknown distributions $(\mathcal{H},\mathcal{B})$, and inherent intractability of $\mathcal{F}$, we frame this as a \textit{learning} problem defined on a parameterized function space completely characterized by the learnable parameters $\Theta$. 
Any such function $f_{\Theta}(\mathbf{b}_{t-1},\mathbf{H}_t)$ generates time-varying power allocations as $\mathbf{p}_t = f_{\Theta}(\mathbf{b}_{t-1},\mathbf{H}_t) = \pi_{\theta}(\mathbf{b}_{t-1}, \mathbf{H}_t;g_{\phi}(\mathbf{H}_t))$ where $\Theta = \{\theta,\phi\}$.
Figure~\ref{F:block} presents an illustration of the proposed model.

Owing to the action-feedback structure of the upper-level problem where the battery state at each time-step is affected by the choice of power allocation, we frame it as a sequential decision-making problem and employ CRL to train $\pi_{\theta}$.
On the other hand, $g_{\phi}$ is trained separately in an unsupervised setting prior to training $\pi_{\theta}$  and the learned weights $\phi$ are frozen such that $g_{\phi}$ essentially serves as a pre-trained feedforward model while training $\pi_{\theta}$. More details on the choice of $g_{\phi}$ are provided in Remark~\ref{R:lower_choice}.\\

\noindent\textbf{Non-myopic Power Allocation} (NMPA) formalizes the upper-level problem as a finite-horizon discounted Markov decision process (MDP) characterized by the tuple $(\mathcal{S},\mathcal{A},\mathcal{P},r,\rho_0,\gamma)$. 
The state space $\mathcal{S}$ and action space $\mathcal{A}$ are continuous. 
Tuples $(\mathbf{b}_{t-1},\mathbf{H}_t)$ for all $t$ constitute $\mathcal{S}$ while the corresponding actions $\mathbf{p}_t$ constitute $\mathcal{A}$.
Since the current $\mathbf{p}_t$ depends only on the last battery state $\mathbf{b}_{t-1}$ and the current channel $\mathbf{H}_t$ but not on the entire history, the Markov assumption~\cite{sutton2018reinforcement} holds. 
The environment is stochastic -- on account of the randomness in the channel states at each time step -- defined by a transition probability distribution $\mathcal{P}: \mathcal{S}\times\mathcal{A}\times\mathcal{S} \to \mathbb{R}$. 
The initial state distribution $\rho_0$ consists of the tuple $(\mathcal{H},\mathcal{B})$ and $\gamma$ is the discount factor that controls the level of myopia in the system. 
A higher $\gamma$ corresponds to a more non-myopic policy. 
A policy is a mapping from $\mathcal{S}$ to $\mathcal{A}$ that aims to optimize a certain objective. 
We redefine the upper-level learner $\pi_{\theta}$ as a parameterized deterministic policy given by 
\begin{align}
    \mathbf{p}_t &= \pi_{\theta}(\mathbf{b}_{t-1},\mathbf{H}_t;\bar{\mathbf{p}}_t) = \mu_{\theta}(\mathbf{b}_{t-1},\mathbf{H}_t)\odot \bar{\mathbf{p}}_t
\end{align}
where $\odot$ is an element-wise multiplication and $\mu_{\theta}: \mathcal{S} \to [0,1]^M$ is a $2$-layered, $3$-filter graph convolutional neural network (GCNN)~\cite{gama2018convolutional} architecture used to learn a vector of scales as
\begin{align}\label{Eq:gcnn}
    \mu_{\theta}(\mathbf{b}_{t-1},\mathbf{H}_t) &= \mathrm{sigmoid} \bigg( \sum_{v=0}^{2} \mathbf{H}_t^v \, \mathbf{Z} \, \theta_{1v} \bigg)\\
    \text{where,}\,\, \mathbf{Z} &= \mathrm{leakyReLU} \bigg( \sum_{v=0}^{2} \mathbf{H}_t^v \, \mathbf{b}_{t-1} \, \theta_{0v} \bigg) \nonumber
\end{align}
with $\theta={\{\theta_{0v},\theta_{1v}\}_{v=0}^2}$.
The scale assigns one scalar weight per transmitter, which modulates $\bar{\mathbf{p}}_t$ based on the current battery condition. 
The primary motivation behind modeling the scale using a GCNN lies in the structure of $\mathbf{H}$, which can be interpreted as the weighted adjacency matrix of a directed graph with self-loops. 
The nodes of this graph are the transceivers. 
The self-loops represent transmission channels, and the remaining edges represent interference.
Moreover, the transmitter-specific elements, such as battery and power allocation, can be considered as signals supported on the nodes of this graph.
Due to this inherent graph structure, a GCNN can learn rich representations for each node by leveraging their local neighborhood information~\cite{chowdhury2021unfolding,schynol2023coordinated}.
Finally, note that the sigmoid non-linearity in~\eqref{Eq:gcnn} restricts the scale within the interval $[0,1]$ element-wise. As, $\bar{\mathbf{p}}_t$ already satisfies~\eqref{Eq:cons1} for all $t$, the product $\mathbf{p}_t$ satisfies~\eqref{Eq:cons1} for all $t$ by construction.

Further, we define a composite reward function $r: \mathcal{S}\times\mathcal{A}\to \mathbb{R}$ as the following:
\begin{align}
    r(\mathbf{b}_{t-1},\mathbf{H}_t;\mathbf{p}_t) = \sum_{i=1}^M c_{i}(\hat{\mathbf{p}}_t, \mathbf{H}_t) - L\mathbbm{1}( [\mathbf{p}_t]_i > [\mathbf{b}_{t-1} - \alpha]_i)
\end{align}
where $\hat{\mathbf{p}}_t =\min(\mathbf{p}_t,[\mathbf{b}_{t-1}-\alpha]_+)$ and $L$ controls the violation penalty at each transmitter. 
Note that the sum-rate component of the reward is computed based on transmission power, which depends on the available battery in addition to the power allocated by the policy, while the penalty component depends only on the allocated power.
For example, if a certain transmitter has its battery level close to depletion and the policy allocates a high power to it at a given time step, it cannot transmit due to lack of battery and therefore cannot contribute to the overall sum-rate. However, a violation penalty is still attributed to the policy as a high power was allocated while the battery was low.
Intuitively, such a scheme has the implicit effect of promoting non-myopic behavior by forcing the policy to be generally conservative in terms of power allocation within an episode and then specifically guiding it to allocate power optimally only when the achievable sum-rate is high.

The overall RL objective can now be defined as:
\begin{align}
    \max_{\theta}&\,\,\mathbb{E}_{\mathcal{H},\mathcal{B}} \Bigg[\sum_{t=1}^T \gamma^{t-1} r\bigg(\mathbf{b}_{t-1},\mathbf{H}_t;\pi_{\theta}(\mathbf{b}_{t-1},\mathbf{H}_t;\bar{\mathbf{p}}_t)\bigg) \Bigg].~\label{Eq:rlopt}
\end{align}
We employ TD3~\cite{fujimoto2018addressing}, a model-free off-policy deterministic policy gradient algorithm specialized for continuous action spaces, to solve~\eqref{Eq:rlopt}.
Its main advantage lies in employing two critic networks to mitigate the value-overestimation issue that plagues actor-critic algorithms. Similar to the policy network, we use GCNNs to model the critics to allow them to leverage the graph structure in $\mathbf{H}$.

As a direct consequence of using GCNNs to model the policy, the trained model can also be deployed in a fully distributed fashion with each transmitter provided with a copy of the trained policy weights along with some additional feedback links~\cite{shi2011iteratively,chowdhury2021unfolding}. However, training has to be centralized under the current formulation. 
Further, the computational complexity of the NMPA framework at inference is given as $\mathcal{O}(M^2D)$, equivalent to that of a GCNN~\cite{kipf2016semi,gama2018convolutional}, where $M$ is the number of transmitters and $D$ is the hidden dimension.

\begin{remark}[Choice of lower-level learner]\label{R:lower_choice}
	\normalfont 
We choose the UWMMSE~\cite{chowdhury2021unfolding} model to serve as $g_{\phi}$. 
UWMMSE has been shown to achieve superior sum-rate performance on instantaneous CSI matrices with reasonable robustness while incurring very low time and computational cost~\cite{chowdhury2021unfolding,chowdhury2022stability,chowdhury2023deep}.
Further, UWMMSE enforces~\eqref{Eq:cons} explicitly as a non-linearity in its architecture. 
The parameters $\phi$ are trained using gradient feedback by maximizing the sum-rate objective~\eqref{E:optimization_problem} over a sufficiently large set of CSI matrices $\{\mathbf{H}_k\}_{k=1}^{N}$ sampled from $\mathcal{H}$. 
It is important to note here that while we present a specific choice of $g_{\phi}$ in this work for the sake of completeness, the proposed NMPA framework is compatible with any algorithm, learnable or otherwise, that provides a solution to~\eqref{E:optimization_problem}-\eqref{Eq:cons}. 
\end{remark}

\section{Numerical Experiments}\label{S:num_exp}

\begin{figure*}
	\centering
	\subfigure[]{
			\includegraphics[width=0.31\textwidth, height=0.23\textwidth]{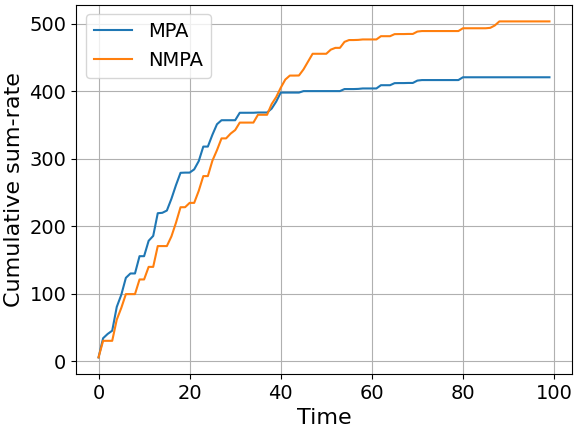}
			\label{Fig:cumulative}
		}	
	\subfigure[]{
			\centering
			\includegraphics[width=0.32\textwidth, height=0.23\textwidth]{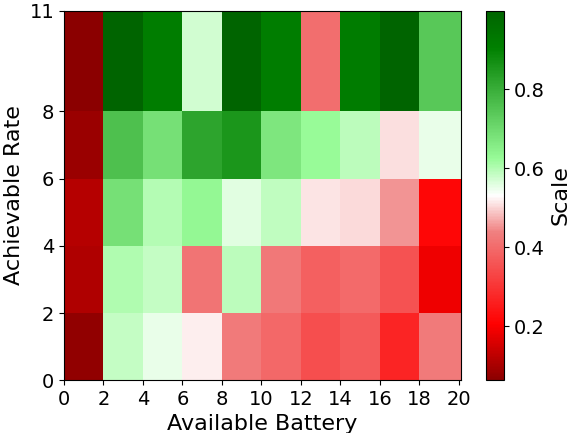}
			\label{Fig:hist}
		}	
	\subfigure[]{
			\includegraphics[width=0.31\textwidth, height=0.23\textwidth]{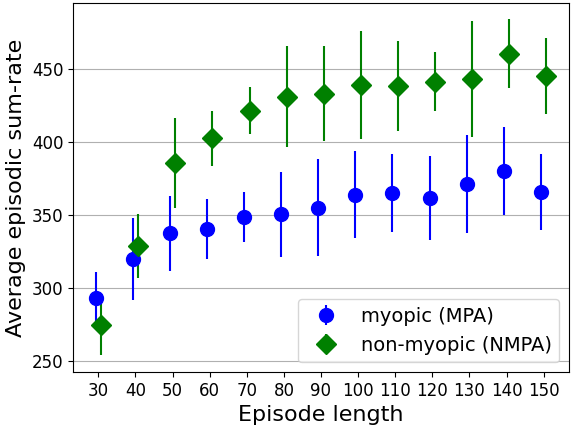}
			\label{Fig:generalize}
		}
		\caption{\small { 
		(a)~Episodic performance of NMPA and MPA in terms of cumulative sum-rate over a randomly sampled episode of length $100$.
            (b)~2D histogram of proportional scale values $\mu_\theta(\cdot)$ generated by NMPA for $10k$ transmitters under multiple configurations of available battery and achievable rate.
		(c)~Generalization performance of NMPA across variable episode lengths. 
            }}
		\label{Fig:results}
\end{figure*}

We now empirically evaluate the performance of the proposed model in terms of average episodic sum-rate and average episodic violations per transmitter under a variable battery budget.
We further demonstrate the generalization performance of the model across various episode lengths.
Maximum instantaneous power is set to $P_{\max} = 1$ unit and maximum battery is set at $B_{\max} = 20$ units.
The violation control parameter is fixed at $L=1$.
Fixed cost $\alpha$ is set to $0.5P_{\max}$.
To model $g_{\phi}$, we use a $4$-layered UWMMSE~\cite{chowdhury2021unfolding} model.
It is trained for a maximum of $10,\!000$ epochs with early stopping. 
The batch size is set at $32$.
The hidden layer dimension for all GCNNs in policy $\pi_{\theta}$ and corresponding critics is set to $32$.
For training $\pi_{\theta}$, we use a replay buffer of size $100k$ transitions with an off-policy batch size of $32$. 
The learning rate for the actor network is set to \num{5e-4} while that for the critic networks is \num{1e-3}. 
The update ratio for target networks is set to \num{1e-3}. 
Training is performed for a maximum of $10,\!000$ episodes of fixed length $100$, with early stopping. 
Inference performances are averaged over $10$ independently sampled episodes.
All computations are performed on an AMD Ryzen Threadripper 3970X 32-Core CPU with 128GB RAM.\footnote{Code to replicate the numerical experiments presented here can be found at \href{https://github.com/archo48/nmpa.git}{https://github.com/archo48/nmpa.git}.}

\vspace{2mm}
\noindent\textbf{Dataset}. We set the wireless network size $M=10$, spatial size $S=60$ units, and transmitter range $R=20$ units for all experiments.
The path loss between transmitter $i$ and receiver $r(j)$ is computed as a function of their corresponding physical distance $d_{ij}(t)$. 
Incorporating small-scale Rayleigh fading, the elements of the channel matrix $[\mathbf{H}_t]_{ij}$ are given by
\begin{align}\label{Eq:topology}
    [\mathbf{H}_t]_{ij} = \frac{1}{1 + d_{ij}^2(t)}\left|\frac{\mathcal{N}(0,1)}{\sqrt{2}} + \text{i}\frac{\mathcal{N}(0,1)}{\sqrt{2}}\right| \quad \forall\,\, i,j.
\end{align}
Since our aim is to emphasize the selectivity of our model in transmitting under ``good" versus ``poor" channel conditions, we manually construct a set of two topologies, low-interference (good) -- wherein intra-transceiver spacing is small while inter-transceiver spacing is large -- and high interference (poor), which is the reverse.
At each time-step $t$, we sample a topology uniformly at random from this set and combine the corresponding path losses with fading effects in~\eqref{Eq:topology}.
The battery budget is sampled from an $M$-dimensional uniform distribution $\mathcal{U}^M_{[0.5B_{\max},B_{\max}]}$.\\ 

\vspace{2mm}
\noindent\textbf{Episodic Behavior}. To evaluate the effectiveness of the proposed NMPA framework, we compare its achieved episodic sum-rate with that of the myopic power allocation (MPA) model, i.e., UWMMSE with battery-agnostic power allocation. 
As shown in Figure~\ref{Fig:cumulative}, the cumulative episodic sum-rate achieved by NMPA at the end of a randomly sampled episode of length $100$ is significantly higher than that of MPA. 
The average improvement in episodic sum-rate achieved by NMPA over MPA, computed over $10$ randomly sampled episodes, is $\sim {15-20\%}$. 
Moreover, this performance improvement is achieved by incurring merely ${0.0005} - {0.0008}$ average violations per transmitter, computed for a $10$-transceiver network over $10$ independent episodes of length $100$.
Clearly, the proposed model learns not to allocate power once the battery is (close to being) depleted. 
And, in doing that, it learns to spend battery intelligently only when the achievable sum-rate is high (good channel conditions).
This behavior can be observed in Figure~\ref{Fig:cumulative}, where the NMPA curve flattens at several time steps initially, for which the MPA model shows an increment, indicating that NMPA chooses not to transmit for those channels, presumably due to their relatively poor channel conditions.
By contrast, MPA spends battery greedily and runs out of battery faster, not being able to leverage good channels occurring later in the episode.
This is further emphasized in Figure~\ref{Fig:hist}, wherein the scale values $\mu_\theta(\cdot)$ for NMPA are high for transmitters that have better channel conditions (high achievable rate) when the available battery is non-zero.
It can be clearly observed that the scales are minimized when the batteries are fully depleted or close to depletion.
The inference time for the proposed hierarchical framework is $7$ ms, wherein the NMPA model takes ${3}$ ms while the remaining time is consumed by the lower-level model.
This fast power allocation is well-suited for rapidly varying channel conditions.

\vspace{2mm}
\noindent\textbf{Generalization performance}. We evaluate the generalization behavior of the NMPA framework by varying the inference episode length in the range $[30,150]$ while the model is trained on a fixed episode length of $100$.
We observe that the average episodic sum-rate achieved by NMPA is consistently higher than that of MPA for episode lengths both greater and smaller than $100$.
This is illustrated in Figure~\ref{Fig:generalize}.
It is important to note that the length of the episode is not known to the model apriori.
Clearly, the model learns to be agnostic to the episode lengths and makes decisions purely based on the immediate state of the system.

\section{Conclusions}\label{S:conc}

We proposed a constrained-reinforcement-learning based non-myopic power allocation method for episodic power allocation under battery constraints.
We employ GCNNs to leverage the underlying graph structure in wireless networks.
The proposed framework is fast, effective, and generalizes across multiple episode lengths. 
It is also compatible with a large class of instantaneous power allocation algorithms.
Future work includes power allocation under multiple time-coupled constraints, distributed training, and applying this framework to mobile wireless networks.\\

\bibliographystyle{IEEEtran}
\bibliography{refs}

\end{document}

%% file: my_sections.tex
\usepackage{needspace}



%% file: figures/tikz_styles.tex
\tikzstyle{phantom vertex} = [ ellipse, 
                               anchor = center, 
                               minimum height = 1*\unit, 
                               minimum width  = 1*\unit,
                               inner sep=0pt,
                               anchor=center]
\tikzstyle{red vertex}   = [black, fill = red!20,   phantom vertex, draw]
\tikzstyle{black vertex} = [black, fill = black!20, phantom vertex, draw]
\tikzstyle{blue vertex}  = [black, fill = blue!20,  phantom vertex, draw]
\tikzstyle{green vertex} = [black, fill = green!20,  phantom vertex, draw]
\tikzstyle{yellow vertex} = [black, fill = yellow!20,  phantom vertex, draw]
\tikzstyle{cyan vertex} = [black, fill = cyan!20,  phantom vertex, draw]
\tikzstyle{vertex}       = [draw, phantom vertex]

\tikzstyle{point} = [ellipse, inner sep=0pt, draw, fill=white, anchor = center,
                     minimum height = 0.05*\unit, minimum width  = 0.05*\unit]